\documentclass[twocolumn,prl,showpacs,nofootinbib]{revtex4}
\usepackage{graphics,graphicx}

\begin{document}
\title{Memory Is Relevant In The Symmetric Phase Of The Minority Game}
\author{K.~H. Ho}
\author{W.~C. Man}
\author{F.~K. Chow}
\author{H.~F. Chau}
\email{hfchau@hkusua.hku.hk}
\affiliation{Department of Physics, University of Hong Kong, Pokfulam Road,
 Hong Kong}
\affiliation{Center of Theoretical and Computational Physics, University of
 Hong Kong, Pokfulam Road, Hong Kong}
\date{\today}

\begin{abstract}
 Minority game is a simple-mined econophysical model capturing the cooperative
 behavior among selfish players. Previous investigations, which were based on
 numerical simulations up to about 100 players for a certain parameter $\alpha$
 in the range $0.1 \lesssim \alpha \lesssim 1$, suggested that memory is
 irrelevant to the cooperative behavior of the minority game in the so-called
 symmetric phase. Here using a large scale numerical simulation up to about
 3000 players in the parameter range $0.01 \lesssim \alpha \lesssim 1$, we show
 that the mean variance of the attendance in the minority game actually depends
 on the memory in the symmetric phase. We explain such dependence in the
 framework of crowd-anticrowd theory. Our findings conclude that one should not
 overlook the feedback mechanism buried under the correlation in the history
 time series in the study of minority game.
\end{abstract}

\maketitle
\section{Introduction}
 Minority game (MG) \cite{MG,Euronews} is the most studied econophysical model
 capturing the minority seeking behavior of independent selfish players. MG is
 a repeated game of $N$ players, each picking one out of two alternatives
 independently in every time step based on the publicly posted minority choices
 of the previous $M$ turns. Those correctly picking the minority choice are
 awarded one mark while the others are deducted one. The aim of each player is
 to maximize his/her own mark. To help players making their choices, each of
 them are assigned once and for all $S$ deterministic strategies. Here, each
 strategy is a map from the set of all possible minority choices of the
 previous $M$ turns, which we call the history, to the set of the two
 alternatives. The performance of each strategy is evaluated according to its
 virtual score, which is defined as the hypothetical mark it would got if the
 strategy were used throughout the game. Among the $S$ assigned strategies,
 every player follows the suggestion of the one with the highest current
 virtual score to choose an alternative. (In case of a tie, a player randomly
 uses one of his/her current best working strategies.) \cite{MG,MG1}

 The number of possible strategies in MG equals $2^{2^M}$. Nonetheless, out of
 these $2^{2^M}$ strategies, only $2^{M+1}$ of them are significantly
 different. These significantly different strategies formed the so-called
 reduced strategy space. In fact, for a fixed $S$ and up to first order
 approximation, the dynamics of MG is robust with $\alpha$, which is the ratio
 of the reduced strategy space size $2^{M+1}$ to the number of strategies at
 play $N S$. \cite{MG1,Euronews}

 The attendance of an alternative $A(t)$ at turn $t$ is defined as the number
 of players choosing that alternative in that turn. Since there is no prior
 bias in choosing the two alternatives in MG, $\left< A(t) \right>_{t,\xi} =
 N/2$ where the average is taken over time $t$ and initially assigned
 strategies $\xi$. In contrast, the variance of attendance per player averaged
 over initially assigned strategies, $\left< \sigma^2 (A(t)) \right>_\xi / N$,
 is a more instructive quantity to study. (We write $\left< \sigma^2 (A(t))
 \right>_\xi$ as $\sigma^2$ and call it simply the variance of attendance from
 now on when confusion is not possible.) Numerical simulation shows that, over
 a wide range of parameters, the $\sigma^2 / N$ against $\alpha$ curve is lower
 than the value in which all players make their choices randomly. This implies
 that these selfish and independent players maximize their own marks
 cooperatively. More importantly, a cusp is found at $\alpha = \alpha_c$; this
 second order phase transition point $\alpha_c$ divides the parameter space
 into the so-called symmetric ($\alpha < \alpha_c$) and asymmetric ($\alpha >
 \alpha_c$) phases \cite{phase}.

 The first numerical study on the effect of history concerning the dynamics and
 cooperative behavior of MG was carried out by Cavagna using $N = 101$, who
 suggested that memory is irrelevant in MG. That is to say, the $\sigma^2 / N$
 against $\alpha$ curve is unaltered if we replace the history in each turn
 with a randomly and independently generated $M$-bit string, while keeping the
 virtual score calculation method unchanged \cite{Cavagna}. (From now on, we
 denote the original MG and the one played using random history strings by
 MG$_\textrm{\scriptsize real}$ and MG$_\textrm{\scriptsize rand}$
 respectively.) Later on, by exploring a wider range of parameters in their
 numerical simulations, Challet and Marsili pointed out that history is
 irrelevant if and only if $\alpha < \alpha_c$. As long as the occurrence of
 history is uniform, $\sigma^2 / N$ should not change when we replace the
 history with a randomly and independently generated time series of $M$-bit
 string. \cite{Memory} Furthermore, Lee argued that although the time series of
 the attendance shows a strong periodic signal, this signal does not affect the
 volatility of the attendance \cite{Lee}. To summarize, the current
 understanding is that memory plays no role on the volatility in the symmetric
 phase of MG.

 In this paper, we perform a large scale numerical simulation of MG played
 using real and random histories with up to $3232$ players for $0.01 \lesssim
 \alpha \lesssim 10$. Our simulation results show that memory is in fact
 relevant in the symmetric phase whenever $\alpha \lesssim 0.2$. Specifically,
 we discover that in such a low value of $\alpha$, the $\sigma^2 / N$ against
 $\alpha$ curves corresponding to the games played using real and random
 histories split. We explain this split by crowd-anticrowd theory developed by
 Hart \emph{et al.} \cite{Crowd1,Crowd2} and argue that finite size effect
 prevents earlier investigations from revealing this discrepancy. Finally, we
 report two scaling relations in the symmetric phase of MG and explain their
 origin.

\section{Numerical Results}
 All the $\sigma^2$'s reported in our simulation are averaged over 1000
 independent runs. And in each run, the variance of attendance is computed from
 the attendance of $15000$ consecutive turns after the system equilibrates.
 Furthermore, we looked at the attendance time series over at least $8\times
 10^6$ iterations in about $50$ independent runs with various values of
 $\alpha$ to verify that the system has indeed equilibrated. Our computation
 requires about $7$~Gflops~yr of instructions.

\begin{figure}[t]
 \begin{center}
  \includegraphics*[scale=0.31]{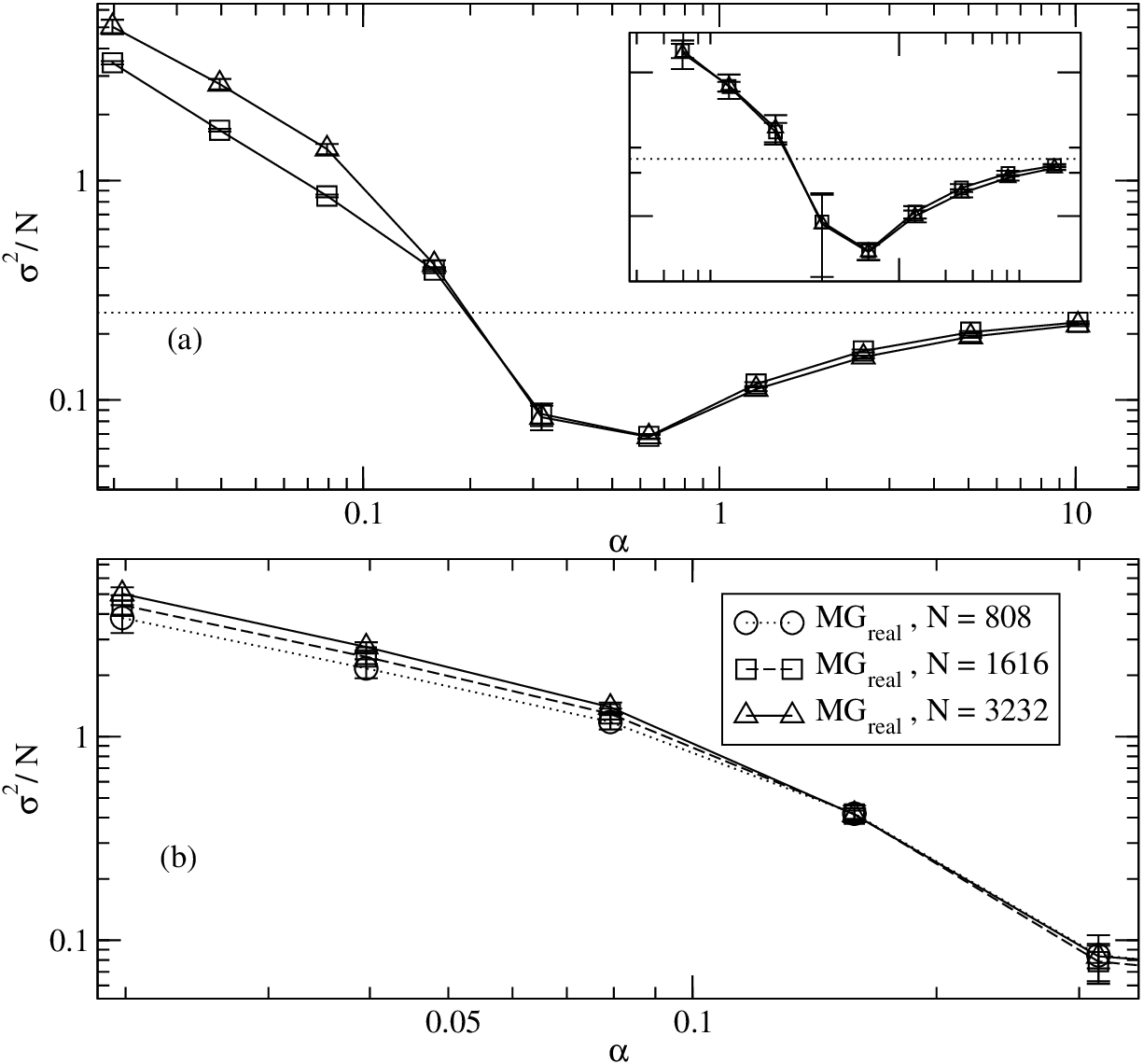}
 \end{center}
 \caption{A plot of $\sigma^2 / N$ against $\alpha$ by varying $M$ with $S=2$.
  Fig.~\ref{F:variance}a compares the dependence of history for $N = 101$
  (insert) and $3232$ (main plot). We represent the
  MG$_\textrm{\scriptsize real}$ and MG$_\textrm{\scriptsize rand}$ data by
  triangles and squares respectively. Fig.~\ref{F:variance}b shows the
  dependence of $\sigma^2 / N$ on $N$ for MG$_\textrm{\scriptsize real}$.}
 \label{F:variance}
\end{figure}

 In Fig.~\ref{F:variance}, we plot $\sigma^2 / N$ against $\alpha$ for
 different values of $N$ for both the MG played using real and random
 histories. By putting $N = 101$, we successfully reproduce the two $\sigma^2 /
 N$ against $\alpha$ curves for MG$_\textrm{\scriptsize real}$ and
 MG$_\textrm{\scriptsize rand}$ obtained by Challet and Zhang in
 Ref.~\cite{Memory} in the range of $0.1\lesssim \alpha \lesssim 10$. In fact,
 these two curves coincide in the symmetric phase. Surprisingly, by increasing
 the number of players $N$, $\sigma^2 / N$ for MG$_\textrm{\scriptsize real}$
 is consistently higher than that of MG$_\textrm{\scriptsize rand}$ for $\alpha
 \lesssim 0.2$. Furthermore, Fig.~\ref{F:variance}b shows that for a fixed
 $\alpha \lesssim 0.2$, $\sigma^2 / N$ for MG$_\textrm{\scriptsize real}$
 increases as $N$ increases. In contrast, we find that the $\sigma^2 / N$
 against $\alpha$ curve for MG$_\textrm{\scriptsize rand}$ is independent of
 $N$. Thus, the discrepancy between the two volatilities increases with $N$
 when $\alpha \lesssim 0.2$. (Although the values of $N$ used in all curves
 reported in this paper are in the form $2^k \times 101$ for some integer $k$,
 our numerical simulation shows that the same conclusions are reached to the
 case of odd $N$.)

\begin{figure}[t]
 \begin{center}
  \includegraphics*[scale=0.31]{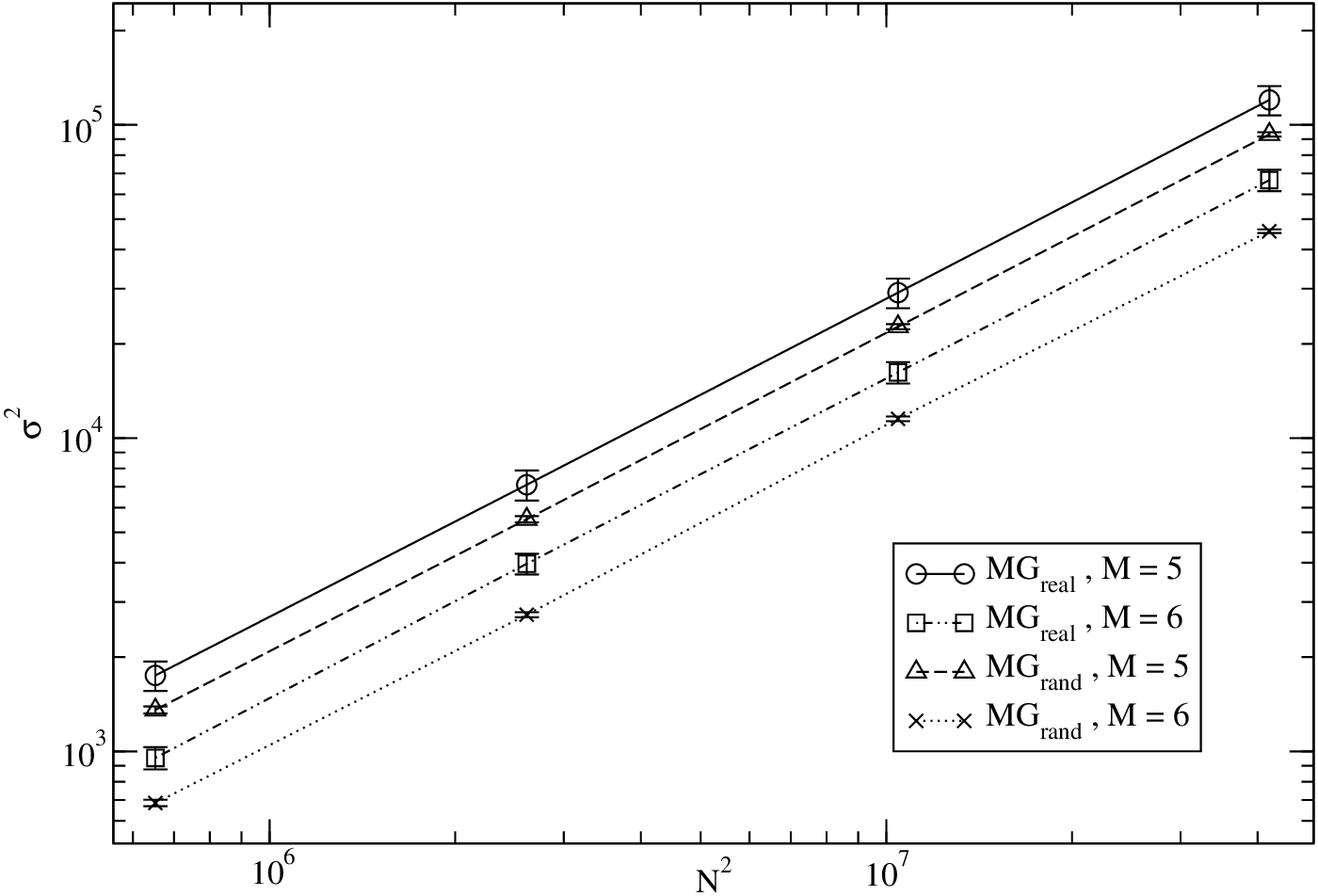}
 \end{center}
 \caption{A plot of $\sigma^2$ against $N^2$. Each curve is drawn by joining
  $\sigma^2$ for different $N^2$ at a fixed $M$ with $S = 2$.}
 \label{F:variance_M}
\end{figure}

 From the above discussions, MG$_\textrm{\scriptsize rand}$ follows the scaling
 relation
\begin{equation}
 \frac{\sigma^2_\textrm{\tiny rand}}{N} \sim f(\alpha,S)
 \label{E:random_scaling}
\end{equation}
 in the symmetric phase ($\alpha < \alpha_c$), where
 $\sigma^2_\textrm{\scriptsize rand}$ is the variance of the attendance for
 MG$_\textrm{\scriptsize rand}$ and $f$ is a scaling function depending only on
 $\alpha$ and $S$. Moreover, Fig.~\ref{F:variance_M} shows that, for a fixed
 memory size $M$, the value of $\sigma^2 / N^2$ is independent of $N$ in the
 symmetric phase. That is to say,
\begin{equation}
 \frac{\sigma^2_\textrm{\tiny real}}{N^2} \sim g_\textrm{\scriptsize real}
 (M,S) > g_\textrm{\scriptsize rand} (M,S) \sim
 \frac{\sigma^2_\textrm{\tiny rand}}{N^2} \label{E:all_scaling}
\end{equation}
 provided that $\alpha < \alpha_c$, where $\sigma^2_\textrm{\scriptsize real}$
 is the variance of attendance for MG$_\textrm{\scriptsize real}$ and $g_i$'s
 are scaling functions depending on $M$ and $S$ only.

\section{The Crowd-Anticrowd Explanation}
 The above numerical findings can be explained by the crowd-anticrowd theory
 proposed by Hart \emph{et al.} using the notion of reduced strategy space
 \cite{Crowd1,Crowd2}. Recall that decisions made by two distinct strategies in
 a reduced strategy space are either mutually anti-correlated or uncorrelated
 when averaged over all possible history strings. Besides, the dynamics is very
 close to the original MG if all strategies are picked from the reduced
 strategy space \cite{MG1}. In this formalism, the crowd-anticrowd theory
 states that every pair of anti-correlated strategies contributes independently
 to the variance of attendance $\sigma^2$ \cite{Crowd1,Crowd2}.

 To understand the difference in volatility between
 MG$_\textrm{\scriptsize real}$ and MG$_\textrm{\scriptsize rand}$, we first
 review the periodic dynamics observed in the symmetric phase of
 MG$_\textrm{\scriptsize real}$
 \cite{Savit,Memory,Crowd1,Crowd2,Manuca,Dynamics}. First, a prominent period
 $2^{M+1}$ peak in the Fourier transform of the minority choice time series for
 MG$_\textrm{\scriptsize real}$ is found. We call this phenomenon ``period
 $2^{M+1}$ dynamics''. In addition, a conspicuous period two peak in the
 Fourier transform of the time series of the minority choice conditioned on an
 arbitrary but fixed history string in MG$_\textrm{\scriptsize real}$ is also
 observed. We refer this as the ``period two dynamics'' in our subsequent
 discussions. Note that the above two dynamics are also observed by replacing
 the minority choice with attendance.

 For MG$_\textrm{\scriptsize real}$ in the symmetric phase, the number of
 strategies at play is much larger that the reduced strategy space size. So, it
 is likely that each strategy in a reduced strategy space is assigned to more
 than one player in this phase. Initially, for a given history string $\mu$,
 every alternative has equal chance to win. And the virtual score of a strategy
 $\beta$ that has correctly predicted the winning alternative is increased by
 one while that of its anti-correlated strategy $\bar\beta$ is decreased by
 one. Since the $M$-bit history string in MG$_\textrm{\scriptsize real}$
 provides complete information of the winning choices in the previous $M$
 turns, players will prefer to use strategy $\beta$ to $\bar\beta$ at the time
 when the same history string appears next. As more players are using the
 strategy $\beta$ in the symmetric phase, these players are less likely to
 guess the minority choice correctly. In MG$_\textrm{\scriptsize real}$, the
 history strings of two consecutive turns are highly correlated. Actually, one
 can convert the history string in turn $t$ to that of turn $(t+1)$ by deleting
 the $(t-1-M)$th turn minority choice from one end of the former string and
 then appending the minority choice in turn $t$ to the other end. Using these
 observations, Challet and Marsili studied the dynamics of
 MG$_\textrm{\scriptsize real}$ by means of a de~Bruijn graph \cite{Memory}. A
 consequence of their analysis is that the history strings in the symmetric
 phase from the $(2^M k + 1)$th to the $[2^M (k+1)]$th turn is likely to form a
 de~Bruijn sequence for any natural number $k$. In other words, for any natural
 number $k$, an arbitrarily given history string $\mu$ is likely to appear
 exactly once between $(2^M k + 1)$th and $[2^M (k+1)]$th turns. Besides, the
 history $\mu$ is likely to appear exactly twice between $(2^{M+1} k + 1)$th
 and $[2^{M+1} (k+1)]$th turns --- one of the turn at which a particular
 alternative wins and the other turn at which the same alternative loses. As a
 result, it is highly probable that the virtual scores of all strategies in the
 $(2^{M+1} k + 1)$th turn agree. Since the decision of a player depends on the
 difference between the virtual scores of his/her strategies, the dynamics of
 the game MG$_\textrm{\scriptsize real}$ has a strong tendency to ``reset''
 itself once every $2^{M+1}$ turns. This is the origin of the period two and
 period $2^{M+1}$ dynamics \cite{Savit,Memory,Crowd1,Crowd2,Manuca,Dynamics}.
 Because of the above two constraints on the history time series, the time
 series of the minority choice as well as that of the virtual score of a
 strategy in MG$_\textrm{\scriptsize real}$ are not random walks.

\begin{figure}[t]
 \begin{center}
  \includegraphics*[scale=0.31]{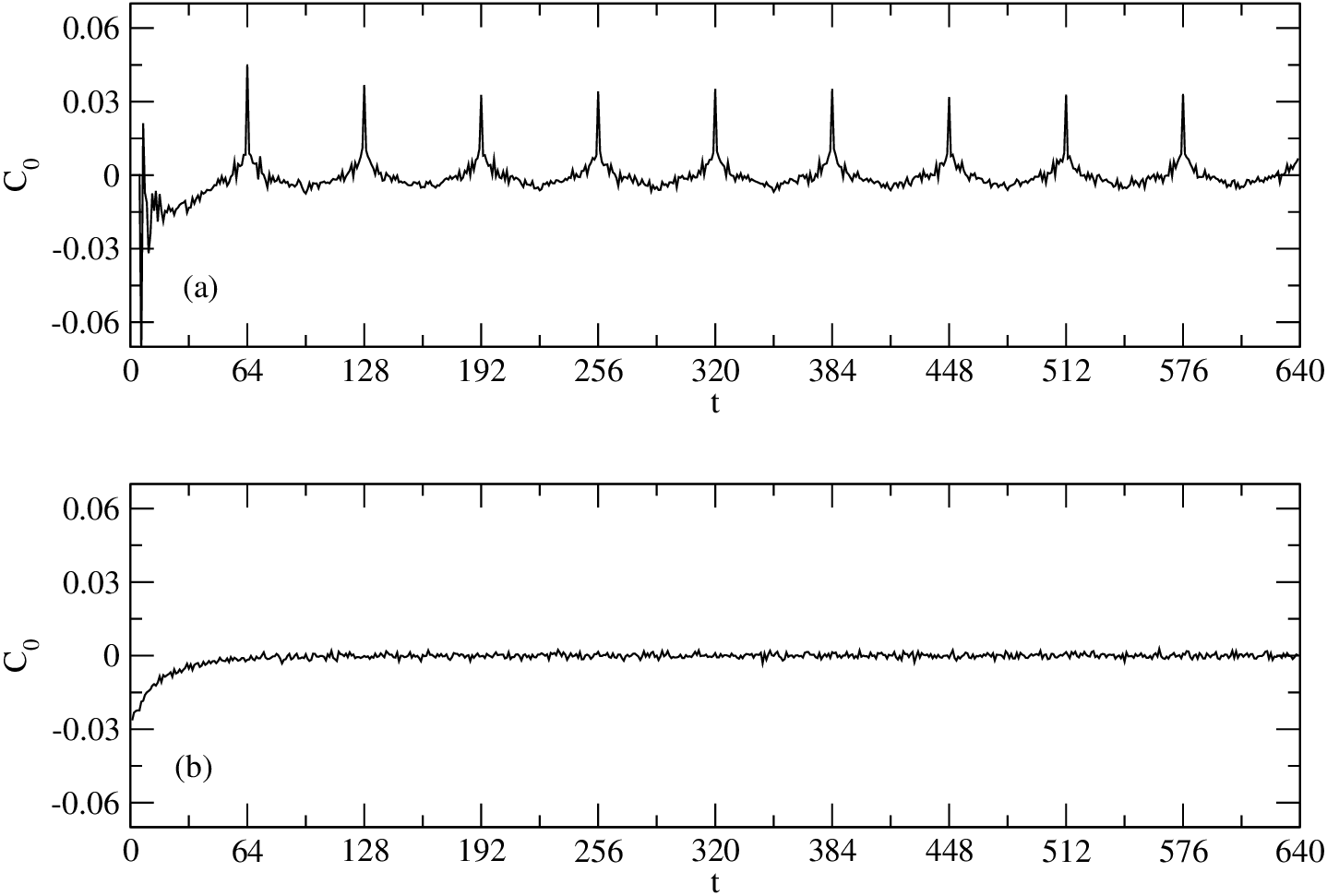}
 \end{center}
 \caption{The auto-correlation function $C_0$ of the minority choice time
  series averaged over 50 runs for (a)~MG$_\textrm{\scriptsize real}$ and
  (b)~MG$_\textrm{\scriptsize rand}$. The parameters used in both plots are $N
  = 1616$, $M = 5$ and $S = 2$.}
 \label{F:period_two_dynamics}
\end{figure}

\begin{figure}[t]
 \begin{center}
  \includegraphics*[scale=0.31]{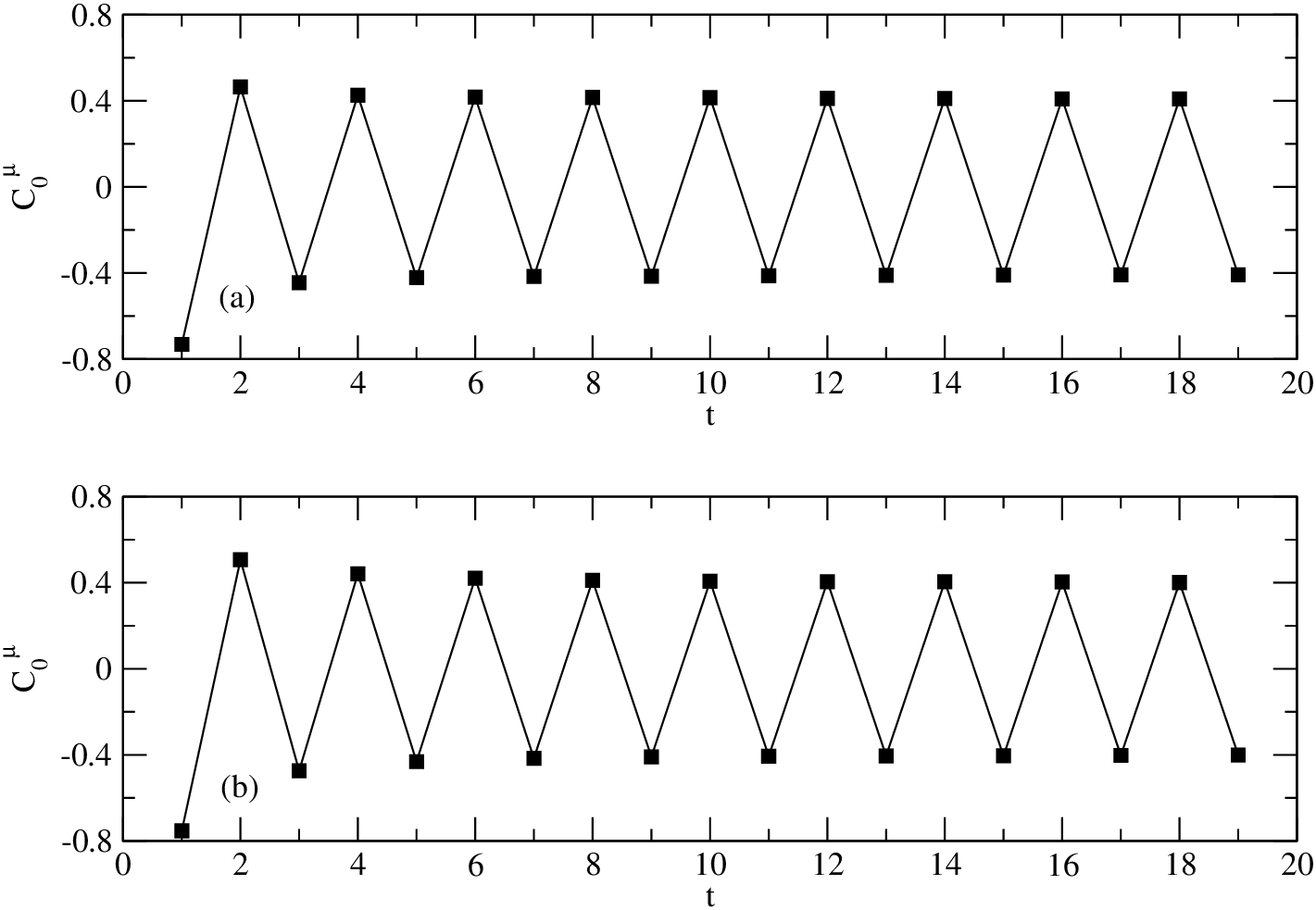}
 \end{center}
 \caption{The auto-correlation function $C_0^\mu$ of the minority choice time
  series conditioned on a given history string $\mu$ averaged over 50 runs for
  (a)~MG$_\textrm{\scriptsize real}$ and
  (b)~MG$_\textrm{\scriptsize rand}$. The parameters used are the same as that
  in Fig.~\ref{F:period_two_dynamics}.}
 \label{F:period_two_dynamics_cond}
\end{figure}

 From the above discussions, we know that the existence of period $2^{M+1}$
 dynamics is closely related to the following three facts that are unique for
 MG$_\textrm{\scriptsize real}$: the $M$-bit history string gives complete
 information on the winning alternatives in the previous $M$ turns, the virtual
 score of a strategy is updated according to the current history string, and
 history strings of two consecutive turns are highly correlated. In contrast,
 for MG$_\textrm{\scriptsize rand}$, the history string does not correlate with
 the winning alternatives, the virtual score is updated according to the
 historical winning alternatives rather than the randomly generated history
 string, and the randomly generated history strings in two distinct turns are
 uncorrelated. Consequently, although MG$_\textrm{\scriptsize rand}$ has a
 uniformly distributed history, it does not have a mechanism to ensure the game
 to ``reset'' itself once every $2^{M+1}$ turns. Thus, there is no reason for
 its history time series to follow a period $2^{M+1}$ dynamics. Indeed, the
 auto-correlation on the minority choice time series in
 Fig.~\ref{F:period_two_dynamics} confirms the absence of period $2^{M+1}$
 dynamics in MG$_\textrm{\scriptsize rand}$ although
 Fig.~\ref{F:period_two_dynamics_cond} shows that
 MG$_\textrm{\scriptsize rand}$ still exhibits period two dynamics. Our finding
 is consistent with Lee's observation that the history occurrence probability
 density function for MG$_\textrm{\scriptsize real}$ and
 MG$_\textrm{\scriptsize rand}$ are different \cite{Lee}. To summarize, for a
 sufficiently small $\alpha$, the entropy of the minority choice time series
 for MG$_\textrm{\scriptsize rand}$ is higher than that of
 MG$_\textrm{\scriptsize real}$.

 We have gathered enough information to explain the discrepancy in the
 $\sigma^2 / N$ against $\alpha$ curves between MG$_\textrm{\scriptsize real}$
 and MG$_\textrm{\scriptsize rand}$. Recall that period two dynamics is
 observed in the symmetric phase of both games. This is because a significant
 number of players are using a particular strategy. And according to the
 crowd-anticrowd theory, this results in a high volatility in the symmetric
 phase in both games \cite{Crowd1,Crowd2}. Further recall that the game
 MG$_\textrm{\scriptsize real}$ is very likely to ``reset'' itself once every
 $2^{M+1}$ turns leading to the period $2^{M+1}$ dynamics in the symmetric
 phase. In contrast, there is no mechanism to ``reset'' the game
 MG$_\textrm{\scriptsize rand}$ once a while for $\alpha \ll \alpha_c$.
 Therefore, the absolute deviation of the virtual score difference between two
 distinct strategies for MG$_\textrm{\scriptsize real}$ is less than that for
 MG$_\textrm{\scriptsize rand}$ whenever $\alpha \ll \alpha_c$. Consequently, a
 player in MG$_\textrm{\scriptsize rand}$ is more likely to stick to a
 strategy. Hence, players in MG$_\textrm{\scriptsize rand}$ cooperate slightly
 better than those in MG$_\textrm{\scriptsize real}$ in the symmetric phase.
 This explains why for any given sufficiently small $\alpha$, the variance of
 attendance per player $\sigma^2 / N$ for MG$_\textrm{\scriptsize real}$ is
 higher than that of MG$_\textrm{\scriptsize rand}$.

 Why do the $\sigma^2 / N$ against $\alpha$ curves for
 MG$_\textrm{\scriptsize real}$ and MG$_\textrm{\scriptsize rand}$ coincide in
 the symmetric phase in the simulations of Cavagna \cite{Cavagna}, Challet and
 Zhang \cite{Memory} together with Lee \cite{Lee}? We believe this is due to
 the finite size effect as they picked $N = 101$ and $S = 2$ in their
 simulations. Due to the small number of players and the small value of $M$
 involved, the fluctuation between different runs overwhelms the discrepancy
 between MG$_\textrm{\scriptsize real}$ and MG$_\textrm{\scriptsize rand}$.

 After discussing the reason why $\sigma^2_\textrm{\scriptsize real}$ and
 $\sigma^2_\textrm{\scriptsize rand}$ differ, we move on to explain the two
 scaling relations (\ref{E:random_scaling})--(\ref{E:all_scaling}). Researchers
 have argued that $\sigma^2$ is well approximated by a function of $\alpha$
 only \cite{MG1,Savit}. Indeed, when $\alpha \gtrsim 0.1$ and hence the number
 of strategies at play is at most ten times the reduced strategy space size,
 the above approximation is reasonably good for MG$_\textrm{\scriptsize real}$
 and MG$_\textrm{\scriptsize rand}$. Nevertheless, our simulation shows that
 this approximation breaks down in MG$_\textrm{\scriptsize real}$ when $\alpha
 \lesssim 0.1$. On the other hand, using mean field approximation, Manuca
 \emph{et al.} showed that
\begin{equation}
 \sigma^2 \approx \frac{N}{4} + \frac{N^2}{3\cdot 2^M} \chi^2 (S)
 \label{E:mean_field}
\end{equation}
 where $\chi$ is a slow varying function of $S$ provided that $\alpha \ll
 \alpha_c$ \cite{Manuca}. Since the correlation between history strings in
 successive turns is ignored in this derivation, the scaling
 relation~(\ref{E:random_scaling}) is applicable to
 MG$_\textrm{\scriptsize rand}$ in the regime of $\alpha \lesssim 0.1$. In
 contrast, our simulation shows that this relation is not applicable to
 MG$_\textrm{\scriptsize real}$ in the same regime.

 Since we have already argued that $\sigma^2_\textrm{\scriptsize real}(M,S) / N
 \geq \sigma^2_\textrm{\scriptsize rand}(M,S) / N$ in the symmetric phase, in
 order to prove the validity of relation~(\ref{E:all_scaling}), it suffices to
 show that $\sigma^2_\textrm{\scriptsize real}(M,S)$ and
 $\sigma^2_\textrm{\scriptsize rand}(M,S)$ both scale as $N^2$. According to
 the crowd-anticrowd theory, for a sufficiently small $\alpha$, all players in
 MG$_\textrm{\scriptsize real}$ hold strategies whose virtual scores are
 similar. Therefore, most players will not stick to a particular strategy when
 making their choices. Besides, the existence of period two dynamics in the
 symmetric phase implies that both $\sigma^2_\textrm{\scriptsize real}$ and
 $\sigma^2_\textrm{\scriptsize rand}$ scales as $N^2$
 \cite{Savit,Crowd1,Crowd2,Manuca}. Thus, relation~(\ref{E:all_scaling}) holds.

 Finally, we briefly explain why $\sigma^2$ decreases as $M$ increases. Since
 the average time between successive appearances of a given history increases
 exponentially with $M$, the absolute deviation of virtual score difference
 between two distinct strategies increases. Therefore, the period two dynamics
 is weakened. So, $\sigma^2$ for MG$_\textrm{\scriptsize real}$ and
 MG$_\textrm{\scriptsize rand}$ decreases as $M$ increases. 

\section{Outlook}
 In this paper, we report that history is relevant in determining the mean
 variance of attendance per player for MG in the symmetric phase when $\alpha
 \lesssim 0.2$ by an extensive numerical simulation using real and random
 histories. We explain our finding using crowd-anticrowd theory. Although all
 graphs shown in this paper are drawn by fixing the number of strategies per
 player $S$ to $2$ and by letting the strategies to be drawn from the full
 strategy space, our conclusions apply equally well to the case of $S>2$ as
 well as the case of drawing strategies from the reduced strategy space. Our
 findings show that the feedback mechanism buried under the correlation in the
 history time series is important in the study of volatility in the minority
 game.

 It is instructive to study the relevance of history in the parameter region of
 $0.2 \lesssim \alpha \leq \alpha_c$. Our numerical simulation suggests that
 memory is irrelevant in this regime and hence the symmetric phase of MG can be
 subdivided into two phases. Further investigation is needed to test our
 hypothesis.

 Finally, it would be nice if one could analytically solve the dynamics of the
 game in the symmetric phase. Since the difference in $\sigma^2 / N$ for
 MG$_\textrm{\scriptsize real}$ and MG$_\textrm{\scriptsize rand}$ in the
 symmetric phase originates from the in period $2^{M+1}$ dynamics, any such
 attempt must take the periodic dynamics of the minority choice time series
 into account --- something that all attempts using replica trick
 \cite{RT1,RT2} and generating functional method \cite{GF1,GF2} so far have not
 been very successful to incorporate.

\par\medskip
\begin{acknowledgments}
\emph{Acknowledgments ---}
 We would like to thank the Computer Center of HKU for their helpful support
 in providing the use of the HPCPOWER~System for the simulation reported in
 this paper. Useful discussions from C.~C. Leung is gratefully acknowledged.
\end{acknowledgments}

\bibliography{mg8.3}
\end{document}